# Multi-Layer Perceptron Neural Network for Improving Detection Performance of Malicious Phishing URLs Without Affecting Other Attack Types Classification

**Pow Cheah Chang**
**Cyber Fellows**
**Tandon School of Engineering**
**New York University, USA**
powchang@nyu.edu

Abstract - The hypothesis here states that neural network algorithms such as Multi-layer Perceptron (MLP) have higher accuracy in differentiating malicious and semi-structured phishing URLs. Compared to classical machine learning algorithms such as Logistic Regression and Multinomial Naïve Bayes, the classical algorithms rely heavily on substantial corpus data training and machine learning experts' domain knowledge to perform complex feature engineering. MLP could perform non-linear separable multi-classes classification and focus less on corpus feature training. In addition, backpropagation weight adjustment could learn which features are more important in differentiating phishing from other attack types.

## I. Introduction

Phishing attack is one of the most significant and prevailing cybersecurity issues. Most of the attacks use sophisticated methods such as deep-fake weblink to trick unknowing users into acquiring sensitive information. Cyberattacks usually use phishing URLs to target financial institutions, government affairs offices, and medical records. In the past decade, the phishing attacks have become increasingly sophisticated and often bluntly mirror the legitimate sites, allowing the attacker to observe everything while the victim is navigating the sites and transverse any additional security boundaries. Riding on the COVID-19 pandemic, phishing becomes the most common attack performed by cybercriminals, with the FBI's Internet Crime Complaint Centre recording over twice as many phishing incidents than any other type of computer crime. [1]

With the proliferation of the Internet of Things and digital transformation driven by the 4th industrial revolution, Uniform Resource Locators (URLs) for resource connectivity and communication networking have grown tremendously. This trend has exposed both the legacy and modern system to the ever-increasing cyber-threats. An attacker can construct web attacks such as SQL, XSS, and information disclosure by embedding executable code or injecting malicious code into the URL. Among the different types of attacks, the data has shown that bad actors leverage phishing URLs to carry out the scam, malware, and ransomware attacks. Furthermore, the forged phishing URL syntax has been evolving rapidly to evade the detection of the blocklist filtering mechanism and make the detection of ever-changing malicious URLs extremely resource-intensive and time-consuming. The hypothesis here is that a lightweight and efficient Multilayer Perceptron Classifier could achieve high effectiveness in detecting phishing URLs and mitigate the risks of cyber-threats.

This paper proposes a deep learning-based method, Multi-Layer Perceptron (MLP), for phishing detection [2]. MPL has higher effectiveness than classical machine learning algorithms such as Logistic Regression and Multinomial Naïve Base, and at the same time could be deployed in edge devices with limited memory size and power storage. Instead of extracting statistical features and lexical features from URLs for those existing standard methods, MLP neural network algorithm focuses on the essential features of the URLs through self-selection, which increases the accuracy to detect subtle differences between genuine and malicious URLs.

This paper uses a dataset from Kaggle [3] to perform the empirical performance comparison between MLP and baseline algorithms. The dataset consists of benign or safe URLs, defacement URLs, phishing URLs, and malware URLs. The experiment runs training without extracting URL features and the statistical features of the links in the web pages. After tokenizing the raw URLs, the tokenized URLs are used for the neural network learning process.

This paper is structured as follows: Section II discusses related works. Section III introduces the proposed phishing detection, and target detection approaches. Section IV shows the experiments to evaluate the proposed approaches and the baselines. Finally, we conclude this work in Section V.

## II. Research related work

In this section, we review some related studies about phishing URLs classification.

### A. Blacklist Approach

The current approach of using traditional spam filtering strategies such as blacklist, and rule-based ranking and Bayesian filters are not effective in detecting new form of phishing URLs which transform and evolve from the old lexical string with slight medication. Furthermore, blacklist repository carries higher cost even though it has high effectiveness for filtering existing and known URLs. The database must be constantly maintained and updated. During the filtering, each of the URLs syntax has to be compared with the inbound URLs, each comparison of the lexical incurs huge overhead and not suitable to be implemented in the real live streaming. Another challenge is that blacklist approach is only effective to certain degree, it is not adaptive enough and has low effectiveness to filter any future new URLs. Once attackers modify the syntax of the URLs, the new URLs could evade the detection easily. Collection of current phishing URLs is another daunting task, a centre repository or webpage could host a universal database for storing phishing URLs, however, this could also be used by the adversary to generate a brand-new URL phishing list for evading the filtering system.

### B. Traditional Machine Learning Models

Traditional machine learning models such as Multinomial Naïve Bayes, Logistic Regression, Random Forests have been widely used in malicious URLs detection. They could achieve high accuracy rate for malicious URLs [4]. However, these models only work well with strong feature engineering pre-processing. The feature extraction and selection require domain knowledge and constant efforts to finetune the model settings. Malicious URLs with virus are easy to spot. They have sketchy syntax, usually executables, which are scannable and blockable. Other types of URLs are links that trigger downloads to the computer. Those links are also easy to identify, and corporate systems can easily block automatic downloads. But phishing URLs, they just look like normal and some even resemble the regular authentic URLs, such as "www.seas.gwu.edu/~mfeldman/". Since phishing URLs look very similar to the website they are impersonating, the current classical machine learning algorithms are not effective enough to filter out those phishing URLs. Many of those URL features have subtle differences comparing to the original authentic URLs, and larger features have to be generated from large dataset for model training and make the deployment at remote site much more challenging.

### C. Deep Learning Approach

Besides the traditional machine learning algorithms, deep learning is an extremely popular approach employed in malware detection due to its high accuracy and capability to perform self-learning based on important features [5]. In deep learning, the feature extraction process is fully automated. As a result, the feature extraction in deep learning is more accurate and result driven. MLPs models are the most basic deep neural network, which is composed of a series of fully connected layers. That is why the proposed solution is using Multi-layer Perceptron which is used to overcome the requirement of high computing power required by modern deep learning architectures such as Convolutional Neural Network (CNN) and Recurrent Neural Network (RNN). MLP is good enough and an optimal solution for simple classification, CNN is good for complicated image classification and RNN is good for sequence processing and these neural networks should be ideally used for the type of problem they are designed for.

## III. Hypothesis and Empirical Work

Attack Types in Testing Dataset Distribution Chart I

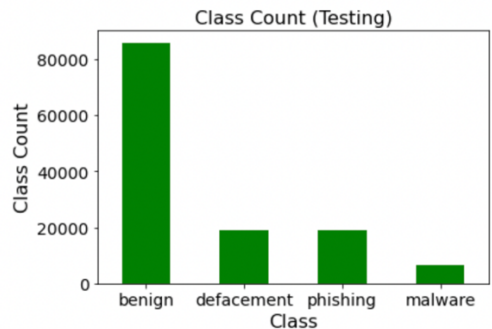

Attack Types in Training Dataset Distribution Chart II

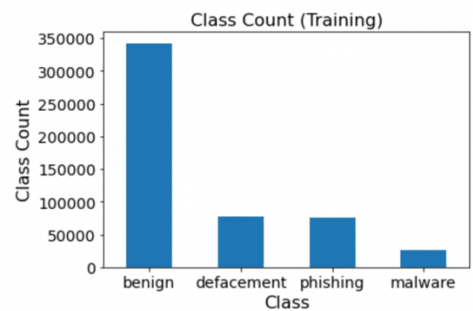

Train and evaluate the Logistic Regression using TF-IDF model Classification Table I

```
              precision    recall  f1-score   support

      benign       0.98      0.94      0.96     90072
  defacement       1.00      0.91      0.95     20859
     malware       0.95      0.99      0.97      6237
    phishing       0.64      0.92      0.75     13071

    accuracy                           0.93    130239
   macro avg       0.89      0.94      0.91    130239
weighted avg       0.95      0.93      0.94    130239
```

Train and evaluate the Logistic Regression using TF-IDF model Confusion Matrix II

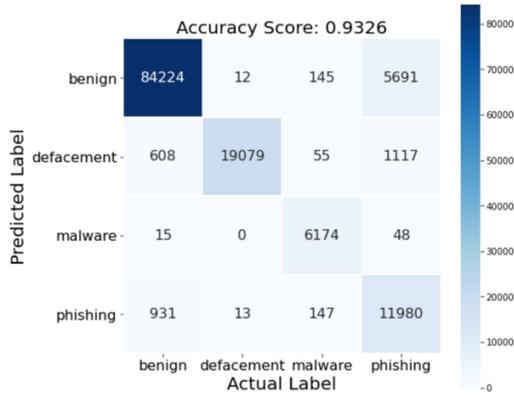

Multi-Layer Perceptron Neural Network Classification Table I

```
              precision    recall  f1-score   support

      benign       0.98      0.97      0.97     86804
  defacement       1.00      0.99      1.00     19203
     malware       0.96      0.98      0.97      6389
    phishing       0.85      0.90      0.87     17843

    accuracy                           0.96    130239
   macro avg       0.95      0.96      0.95    130239
weighted avg       0.96      0.96      0.96    130239
```

Multi-Layer Perceptron Neural Network Classification Table I

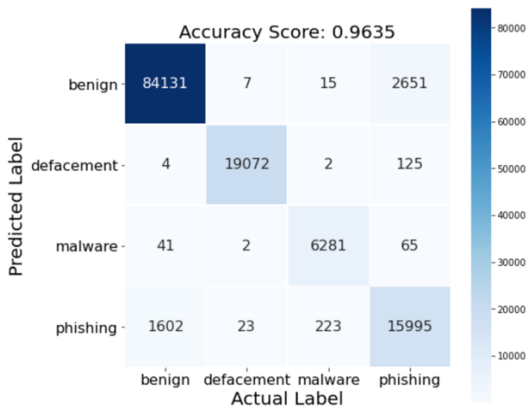

This session presents the empirical comparison of accuracy between Machine Learning Logistic Regression using TF-IDF (term frequency - inverse document frequency) and Multi-Layer Perceptron Neutral Network. As shown in the precision, recall, and accuracy comparison table, other types of attacks such as benign, defacement, malware have comparable performance between the two algorithms. However, there is a significant improvement in using Multi-Layer Perceptron (MLP) Neural Network. The precision of phishing detection improved from 0.64 to 0.85 without using advanced feature engineering for both algorithms to achieve an apple-to-apple comparison. MLP has a more substantial capability to differentiate benign from phishing, and hence from the confusion matrix, it shows almost half reduction of misclassifying benign as phishing. The result indicates that despite the close imitation of phishing URLs to the original version, MLP has a more robust capability to pick up the subtle difference between the two. Empirical run proves the hypothesis to be correct that MLP has performed better than the classical machine learning algorithm in filtering out those phishing URLs that are ever-evolving and strikingly resemble original authentic URLs. The overall accuracy improves from baseline of 0.9326 to 0.9635.

IV. Conclusion And Future Work

In this paper, I propose using light-weight Multi-Layer Perceptron to detect the malicious phishing URLs. Phishing URLs closely imitate the original authentic URLs, this makes it extremely challenging to differentiate the malicious from benign URLs. Feature selection and engineering in classical machine learning may cause degradation of performance in classifying other attack types. The empirical work done using MLP against Logistic Regression shows significant improvement in detecting phishing URLs with low false positive and false negative rate.